# Stimulated Raman adiabatic passage: effect of system parameters on population transfer

Fatemeh Ahmadinouri ,[1] Mehdi Hosseini[1a] and Farrokh Sarreshtedari [2]

[1]*Department of physics, Shiraz University of Technology, Shiraz, 313-71555, Iran*

[2]*Magnetic Resonance Research Laboratory, Department of Physics, University of Tehran, Tehran 143-9955961, Iran*

The Stimulated Raman Adiabatic Passage (STIRAP) procedure is a robust and complete population transfer method which have various application in chemistry and atomic physics. Here, we study the effects of one-photon detuning, transition time, pulse width, and pulse delay parameters on the population transfer via STIRAP and b-STIRAP techniques. Moreover, the impact of the field amplitude which decreases the population transfer has been analyzed and it is shown that in b-STIRAP, the robustness of the complete transition can be improved by decreasing the field amplitude in the single-photon resonance condition.

## I. INTRODUCTION

The population transfer by an adiabatic passage is an important and prominent approach in quantum control [1–4]. Frequency-chirped laser [5–7], *Landau-Zener* technique [8–10], and

---

[a] hosseini@sutech.ac.ir

Stimulated Raman Adiabatic Passage (SRIRAP) [11–17] are some of the important methods of the adiabatic population transfer of the quantum systems.

Among different methods of the population transfer based on the adiabatic passage, STIRAP is a powerful and simple method that because of its robustness and complete population transfer, much attention has been widely paid to this technique [12,14]. The STIRAP is based on a $\Lambda$-linkage pattern which the population is transferred from an initially populated state ($|1\rangle$) to an excited state ($|3\rangle$) via an intermediate state ($|2\rangle$) [14,18]. In this scheme, the intermediate state is coupled to the ground and excited states by two transition dipole moments. This is while the ground and excited states are not directly dipole coupled [19]. In this configuration, the energy of the intermediate state is considered more than the initial and final states. The STIRAP method is designed in such a way that the complete population is adiabatically transmitted to the final state so that no significant population is placed in the intermediate state. Indeed, the adiabatic theorem guarantees the system population is completely trapped to an adiabatic state which is only a linear combination of the initial and final states without any population of the intermediate state. The intermediate state has no population during the interaction. Because of that, this adiabatic state is named *dark state* [3,11,14,18]. In this approach, two delayed laser pulses are utilized wherein states $|1\rangle$ and $|2\rangle$ are coupled via a pump pulse while states $|2\rangle$ and $|3\rangle$ are coupled via a Stokes pulse in such a way that in a counterintuitive pulse sequence, the Stokes pulse precedes the pump pulse while the two pulses are partially overlapped [11,14]. Instead, in an intuitive pulse sequence, the Stokes pulse is applied after the pump pulse and it provides a pathway which the three states are populated and is called b-STIRAP (bright-STIRAP) [16]. In the b-STIRAP method, a significant and efficient population transfer can be occurred if the lifetime of the intermediate state is longer than the transition time [14].

The STIRAP mechanism which was introduced in 1980 by Gaubatz *et al*. [20], has potential properties that led to widespread applications in many fields of chemistry and physics such as quantum information [21–25], superconductors [26,27], and Bose-Einstein condensates [28–33] while nowadays, has been numerically and experimentally studied [12,34,35].

In this work, the effect of the system parameters on the population transfer are extensively investigated for the two approaches of STIRAP and b-STIRAP, using the Jaynes-Cummings (JC) model. Moreover, it is shown that by sufficient decreasing of the field amplitude and adjusting the pulse width and pulse delay, the robustness of the population transfer can be improved in the regions of positive pulse delay (b-STIRAP) while the transition time is very low.

## II. THEORETICAL MODEL

Here, the effect of the two delayed pulses on the population transfer has been studied in a three-level lambda (Λ) configuration. In this scheme, the ground, intermediate, and excited states are denoted by |1⟩, |2⟩, and |3⟩, respectively which the |1⟩ and |2⟩ states are coupled by a pump Rabi frequency ($\Omega_P$) while the |2⟩ and |3⟩ states are connected by a Stokes Rabi frequency ($\Omega_S$). These Rabi frequencies are evaluated with the atomic dipole moment and the field strength. In this work, the full quantum model of the *Jaynes-Cummings* is investigated for investigation of the population dynamics in the interaction of a Λ-type system with an electromagnetic field. The JC Hamiltonian is defined by [36,37]:

$$H = \hbar(\sum_{i=1}^{3} \sigma_{ii}\omega_i + \omega_{L_P}\alpha_P^\dagger\alpha_P + \omega_{L_S}\alpha_S^\dagger\alpha_S + \omega_P(\alpha_P\sigma_{21} + \alpha_P^\dagger\sigma_{12}) + \omega_S(\alpha_S\sigma_{23} + \alpha_S^\dagger\sigma_{32})) \qquad (1)$$

In this Hamiltonian, $\hbar$ is Planck constant, field operators of the pump and Stokes are shown by $\alpha_p$ and $\alpha_s$, $\sigma_{ii} = |i\rangle\langle i|$ are the state numbers, $\sigma_{ij} = |i\rangle\langle j|(i \neq j)$ represents the transition operators, $\omega_i$ is Bohr transition frequency, the pump and Stokes laser frequencies are $\omega_{L_P}$ and $\omega_{L_S}$, and the coupling strength between the pump and Stokes fields and atom are described by $\omega_P$ and $\omega_S$ which are proportional to the Rabi frequencies by [38]:

$$\Omega_{P,S} = \sqrt{(\omega_{L_{P,S}} - \omega_{21,23})^2 + (2\omega_{P,S})^2} \tag{2}$$

Where $\omega_{21}$ and $\omega_{23}$ are the difference between the Bohr transition frequencies and given by $\omega_{21} = \omega_2 - \omega_1, \omega_{23} = \omega_2 - \omega_3$. By acting the JC Hamiltonian (Eq. (1)) on the eigenvectors $|g;n,m-1\rangle, |e;n-1,m-1\rangle, |g;n-1,m\rangle$, which are corresponding to the |1>, |2>, and |3> states in the archetype of STIRAP procedure, Eq. (3) is achieved.

$$H = \hbar \begin{bmatrix} \omega_1 + n\omega_{L_P} + (m-1)\omega_{L_S} & \omega_P\sqrt{n} & 0 \\ \omega_p\sqrt{n} & \omega_2 + (n-1)\omega_{L_P} + (m-1)\omega_{L_S} & \omega_S\sqrt{m} \\ 0 & \omega_S\sqrt{m} & \omega_3 + (n-1)\omega_{L_P} + m\omega_{L_S} \end{bmatrix} \tag{3}$$

In this Hamiltonian, $n$ and $m$ are photon numbers of the pump and Stokes modes, respectively which assuming $n$ and $m$ equal to one, the traditional STIRAP Hamiltonian can be achieved [3,11] (Eq. 4)

$$H = \hbar \begin{bmatrix} 0 & \omega_P & 0 \\ \omega_P & \Delta_P & \omega_S \\ 0 & \omega_S & \delta \end{bmatrix} \tag{4}$$

Where $\Delta_P$ and $\Delta_S$ are the so-called "*one-photon detuning*", namely $\Delta_P = \omega_{21} - \omega_{L_P}, \Delta_S = \omega_{23} - \omega_{L_S}$ and $\delta$ is simplify taken in the form of $\delta = \Delta_P - \Delta_S$ and is named "*two-photon detuning*".

Considering the time-dependent Schrödinger equation and the system Hamiltonian (Eq. (3)), corresponding wave functions are obtained for analyzing of the population transfer. To this end, the transition probabilities are numerically investigated by the *Runge–Kutta* method.

It is important to note that in our calculations, for simplicity we have taken the system as a dimensionless system so that the parameters are defined by:

$$\tilde{\omega}_{23} = 1, \omega_{21} = \tilde{\omega}_{21}\omega_{23}, \omega_P = \tilde{\omega}_P\omega_{23}, \omega_S = \tilde{\omega}_S\omega_{23}, \omega_{L_P} = \tilde{\omega}_{L_P}\omega_{23}, \omega_{L_S} = \tilde{\omega}_{L_S}\omega_{23}, t = \frac{\tilde{t}}{\omega_{23}} \tag{5}$$

In these simulations, the initial conditions emphasize that the system is in the ground state at the beginning of the interaction so that by applying the two delayed pulses, the system population is adiabatically and irreversibly transmitted into the excited state. For this purpose, we have investigated mean values in the 10% Final Probability of Excited state ($|3\rangle$) which is named as FPE in this work.

In this work, using the dimensionless parameters, the pump and Stokes photon numbers are equal to one and the pump transition frequency is assumed to be $\tilde{\omega}_{21} = 2$. The interaction time is also considered in such a way that the stability of the final probabilities is ensured. Moreover, the pump and Stokes pulses are determined by a Gaussian distribution (Eq. (6)) as [39]:

$$\tilde{\omega}_{P,S}(t) = \tilde{\omega}_{\max_{P,S}} \exp\left[\frac{-(\tilde{t} - \tilde{t}_0)^2}{2\tilde{\tau}^2}\right] \tag{6}$$

In this equation, $\tilde{\omega}_{\max_{P,S}}$ is the normalized peak amplitude of the pulses, $\tilde{t}_0$ is the time of the peak amplitude, and $\tilde{\tau}$ determines the pulse width.

### III. RESULTS AND DISCUSSION

For investigating the effect of various system parameters on the population dynamics, the obtained results are presented as FPE figures. Fig. 1(a) shows the FPE versus pulse width and pulse delay for one-photon resonance ($\tilde{\Delta}_{P,S} = 0, \tilde{\omega}_{L_P} = 2, \tilde{\omega}_{L_S} = 1$) and $\tilde{\omega}_{\max_{P,S}} = 1$. In this figure, the pump and Stokes pulse widths are identical and the time at the peak amplitude of the pump pulse is considered to be zero ($\tilde{t}_{0_P} = 0$). With this assumptions, $\tilde{t}_{0_S}$ is a criterion for the pulse delay. In this case, the system population is transferred in the counterintuitive pulse sequence (STIRAP) for the negative pulse delays ($\tilde{t}_{0_S} \prec 0$) while the population transfer occurs in the intuitive pulse sequence (b-STIRAP) for the positive pulse delays ($\tilde{t}_{0_S} \succ 0$). It should be noted that the red regions in the FPE figures express that the complete population transfer has been occurred.

Fig.1 (a) reveals that all of the system population is transferred to the final state in regions of the negative pulse delay (STIRAP) while no robust and complete population transfer to the excited state is occurred in the b-STIRAP areas[17,40]. It can be seen that the increase of the pulse width is necessary for the complete population transfer when the pulse delay increases toward more negative times. We can also see an oscillating behavior in the population evolutions when the two pulses are almost simultaneous.

Fig. 1(b) reveals the FPE versus pulse width and pulse delay for one-photon detuning $\tilde{\Delta}_P = 0.5, \tilde{\Delta}_S = 0$ and $\tilde{\omega}_{\max_{P,S}} = 1$. This figure illustrates that by getting away from the one-photon resonance, the extent of complete transition region reduces [11]. Similar output results is obtained for $\tilde{\Delta}_P = -0.5, \tilde{\Delta}_S = 0$ and $\tilde{\Delta}_P = 0, \tilde{\Delta}_S = \pm 0.5$.

For the detailed analysis of the population evolutions, Fig. 2 is represented in the condition of far off single-photon resonance. The FPE versus Stokes laser frequency and pulse delay for $\tilde{\Delta}_P = 0, \tilde{\tau} = 30$ and $\tilde{\omega}_{\max_{P,S}} = 1$ is depicted in the left-hand part of Fig. 2. It shows that by increasing the pulse delay in the STIRAP areas and moving away from the regions of the full pulse overlap, the complete population transfer occurs in the wide range of the Stokes laser frequency. This rang is maximized in the specified pulse delay while the area extent would be reduced hereafter and eventually, this range has reached zero [41]. This implies that in the optimum delay, the complete transition can be achieved by the maximum one-photon detuning of the Stokes pulse. This figure also reveals that in the case of two-photon resonance ($\tilde{\delta} = 0, \tilde{\omega}_{L_S} = 1$), the complete transition is occurred while extent of the pulse delay is maximized. Moreover, in the regions of positive pulse delay, an oscillating behavior can be seen around one-photon resonance $\tilde{\Delta}_S = 0$ while no robust population transfer has taken place in this area. Our finding indicates similar results for $\tilde{\Delta}_S = 0, \tilde{\tau} = 30$ while the pump laser frequency and the pulse delay has changed. The FPE versus Stokes and pump laser frequencies for $\tilde{t}_{0_S} = -80, \tilde{\tau} = 50$ and $\tilde{\omega}_{\max_{P,S}} = 1$ is shown in the right-hand part of the Fig. 2. This figure emphasizes that the maximum range of the complete transition is achieved for the one-photon resonance ($\tilde{\Delta}_{P,S} = 0, \tilde{\omega}_{L_P} = 2, \tilde{\omega}_{L_S} = 1$) while the increase of the one-photon detuning causes area extent of complete population decreases [11].

As mentioned before, a remarkable feature of the counterintuitive pulse ordering (STIRAP method) is the straight population transfer from the initial state to the final state in such a way that the intermediate state (dark state) remains essentially unpopulated during the process. This implies that the desired transition is independent of the dark state lifetime. The transition probabilities versus time for |2⟩ and |3⟩ states and $\tilde{\Delta}_{P,S} = 0, \tilde{\omega}_{\max_{P,S}} = 1$ are depicted in Fig.3. In this figure, it can be seen that the intermediate level is unpopulated in the all interaction time [11,14].

In the left-hand part of the figure, the transition probabilities versus time for $\tilde{t}_{0_S} = -80$ and different pulse widths are illustrated. The transition probabilities versus time for $\tilde{\tau} = 50$ and different pulse delays are indicated in the right-hand portion of the figure. This figure reveals that by increasing pulse width, the transition time is increased (Fig. 3 (a)) while by decreasing the negative pulse delay, the transition time and the oscillations of the transition probability diagrams are increased (Fig. 3 (b)) [42]. The impact of these oscillations on the population transfer can be observed when the two pulses fully overlap and the delay pulse is nearly equal to zero as shown in Fig. 1 (a). Therefore, the faster complete transition can be achieved by increasing the delay pulse and decreasing the pulse width.

In order to investigate the effect of the difference between the pulse widths on the transition probabilities, the FPE versus Stokes pulse width and pulse delay for $\tilde{\Delta}_{P,S} = 0, \tilde{\tau}_P = 30, \tilde{\omega}_{\max_{P,S}} = 1$ is shown in Fig. 4. This figure indicates when the two pulses approximately overlap, the complete transition happens only when the pulse width are identical. The increase of the pulse delay causes the full population transfer occurs in the regions with the more width difference. This trend continues until a specified pulse delay and afterward, the increase of the Stokes pulse width is necessary when the pulse delay is increased. Moreover, this figure implies when the pulse delay is near zero, the population transfer occurs only where the width of the two pulses is almost equal ($\tilde{\tau}_{S,p} = 30$). The similar conclusion can be obtained when the pump pulse width and the pulse delay vary while $\tilde{\Delta}_{P,S} = 0, \tilde{\tau}_S = 30, \tilde{\omega}_{\max_{P,S}} = 1$.

The effect of the decreasing the field amplitude is investigated in Fig. 5[11,37]. To this end, the FPE versus pulse width and pulse delay is illustrated for single-photon resonance ($\tilde{\Delta}_{P,S} = 0$) and $\tilde{\omega}_{\max_{P,S}} = 0.1$. This figure implies that by reducing the pulse amplitudes, the robustness of the complete population transfer is improved in the b-STIRAP (positive pulse delays) case (comparing with Fig. 1(a)). Moreover, this figure confirms that in the negative pulse delays, the robust population transfer can be achieved for different pulse amplitudes. It also shows that by setting the specified pulse width, the complete transition occurs in the wide range of the positive pulse delay. Our finding shows that by further decreasing the pulse amplitude, the wide area of the full transition increases and robustness of the complete population transfer enhances in the b-STIRAP areas. In other words, the stability of the results augments against small variations of parameters. As a consequence, by decreasing the pulse amplitudes, the robust and complete population transfer can be achieved both in forms STIRAP and b-STIRAP. Although, the three-level of the system are employed for the population transfer to the excited state in the b-STIRAP procedure, the irreversible and complete population transfer can be achieved when the transition time is short compared to the lifetime of the intermediate state (bright state). Using both The STIRAP and b-STIRAP methods helps us to simultaneously transfer the population of two separate states in which may have applications in quantum information [15,16].

In Fig. 5, we have defined six points. The transition probabilities versus time for these points are depicted In Fig. 6. By comparing the a, b and c cases it is obvious that reduction of the pulse width in the b-STIRAP causes a reduction in oscillations of the transition probability [42]. Hence, the bright state is less involved in the population transfer. By comparing the d, e and f cases, it can be seen that by decreasing the positive pulse delay at a determined pulse width, the population oscillations are more concentrated and therefore, the transition time is reduced [11]. As a result, by decreasing the pulse delay and the pulse width, the transition time and the population oscillations are reduced in the areas of the complete transition for the positive pulse delay. In other words, the bright state is interacting with the two other states in less time and accordingly, for excitation of the system population to the excited state with the least decay, the less lifetime of the bright state is needed.

Another approach in the population transfer would be achieved in the positive pulse delay if the parameters are tuned in a far off one-photon resonance [40,43]. For detailed investigation, this approach is studied in figures 7-10. The FPE versus pulse width and pulse delay for $\tilde{\Delta}_{P,S} = 0.5$ and $\tilde{\omega}_{\max_{P,S}} = 1$ is shown in Fig. 7. This figure reveals when the pulse delay is close to zero, the system population, oscillates more clearly compared with the Fig. 1 (a). In this figure, it is also evident that the complete population transfer can be obtained under the situation of far off one-photon resonance in the b-STIRAP method.

The transition probabilities versus time for |2⟩ and |3⟩ states and $\tilde{\Delta}_{P,S} = 0.5, \tilde{\omega}_{\max_{P,S}} = 1$ are illustrated in Fig. 8. In the left-hand side of the figure, the positive pulse delay is assumed $\tilde{t}_{0_S} = 80$ and the transition probabilities for different pulse widths are illustrated. This figure reveals that by increasing the pulse width, the transition time to state |3⟩ is increased. The bright state is more involved in the transition process. This is while, the oscillation probabilities of state |3⟩ is also increased. [42]. In the right-hand side of the Fig. 8, the pulse width is supposed as $\tilde{\tau} = 50$ and the transition probabilities for different pulse delays are indicated. As it can be seen, by increasing the pulse delay, the time of transition to the |3⟩ state is nearly constant and only the transition position is shifted. Moreover, Contrary to STIRAP, the probability oscillations are reduced by increasing the pulse delay. The population transfer has also included the bright state for more time. The implication is that in the b-STIRAP process under the condition of far off one-photon resonance, the bright state is less involved in the transition process by decreasing the pulse width and pulse delay.

The effect of the one- photon detuning value in the b-STIRAP process is investigated in Fig. 9. In this figure, the FPE versus Stokes and pump laser frequencies for $\tilde{t}_{0_S} = 80, \tilde{\tau} = 50$ and $\tilde{\omega}_{\max_{P,S}} = 1$ is shown. It is shows that no complete population transfer occurs in the case of single-photon resonance ($\tilde{\omega}_{L_P} = 2, \tilde{\omega}_{L_S} = 1$) which it is in accordance with our results in Fig. 1 (a). By moving away from the one-photon resonance area, region of the complete transition is increased. This trend continues to a specified limit and afterward, this area is reduced [12].

The transition probabilities versus time for three states, $\tilde{t}_{0_S} = 80, \tilde{\tau} = 50, \tilde{\omega}_{\max_{P,S}} = 1$ and different one-photon detuning are illustrated in Fig. 10. In the left-hand portion of the figure (a), the single-photon detuning is equal to $\tilde{\Delta}_{P,S} = 0.2, \tilde{\omega}_{L_P} = 2.2, \tilde{\omega}_{L_S} = 1.2$ while it is equal to $\tilde{\Delta}_{P,S} = 4, \tilde{\omega}_{L_P} = 6, \tilde{\omega}_{L_S} = 5$ in the right-hand portion of this figure (b). This figure shows that for larger one-photon detuning, the affiliation of the population transfer to the bright state is significantly reduced [11,16,17].

As mentioned, in the positive pulse delay (b-STIRAP), the dependence of the transition to the bright state can be reduced in the far off single-photon resonance condition. However, by reducing the bright state duration, it is possible that the complete population is irreversibly and exactly transferred to the |3⟩ state during interaction even with less lifetime of the bright state.

## IV. CONCLUSION

Here, the impact of the one-photon detuning, transition time, pulse width, and pulse delay parameters on the population transfer is investigated for the positive and negative pulse delay (STIRAP and b-STIRAP). Moreover, in the b-STIRAP method, the transition probabilities have been investigated in two different cases: i) the field strength changes and ii) the far off single-photon resonance. It has been shown that in the case of decreasing the field amplitude, the complete transition is observed in the regions with low pulse width while in the far off single-photon resonance condition no robust transition is achieved.

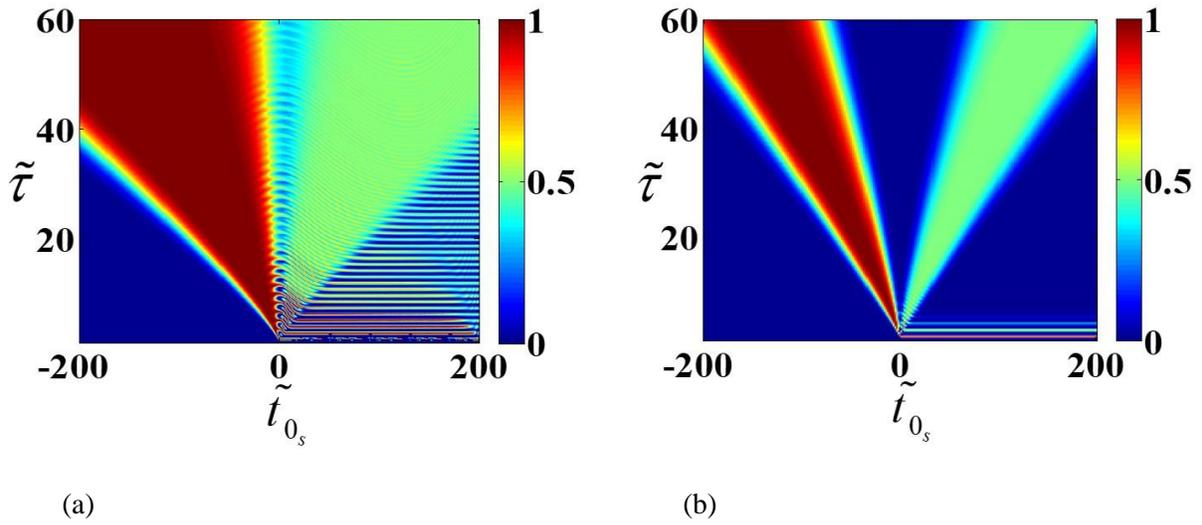

(a)                                        (b)

FIG. 1. FPE versus pulse width and pulse delay for $\tilde{\omega}_{\max_{P,S}} = 1$ and (a) $\tilde{\Delta}_{P,S} = 0$ (b) $\tilde{\Delta}_P = 0.5, \tilde{\Delta}_S = 0$

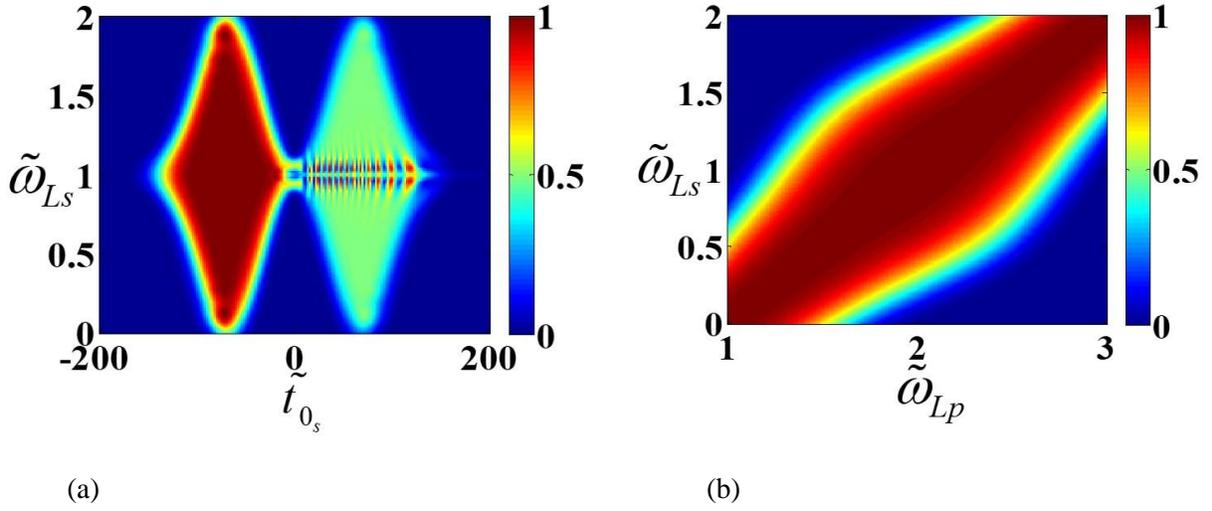

(a)                                    (b)

FIG. 2. FPE versus Stokes laser frequency and (a) pulse delay for $\tilde{\Delta}_P = 0, \tilde{\tau} = 30, \tilde{\omega}_{\max_{P,S}} = 1$

(b) pump laser frequency for $\tilde{t}_{0_s} = -80, \tilde{\tau} = 50, \tilde{\omega}_{\max_{P,S}} = 1$

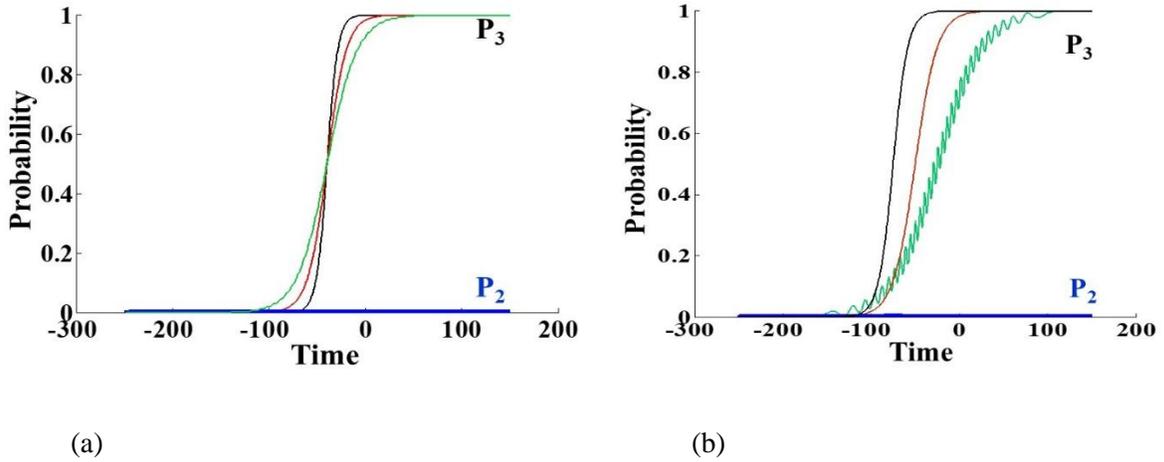

(a)                                    (b)

FIG. 3. Transition probabilities versus time for |2⟩ and |3⟩ states and $\tilde{\Delta}_{P,S} = 0, \tilde{\omega}_{\max_{P,S}} = 1$ (a) $\tilde{t}_{0_s} = -80$

, $\tilde{\tau} = 30$ (black line), $\tilde{\tau} = 40$ (red line), and $\tilde{\tau} = 50$ (green line) (b) $\tilde{\tau} = 50$, $\tilde{t}_{0_s} = -150$ (black line),

$\tilde{t}_{0_s} = -100$ (red line), and $\tilde{t}_{0_s} = -50$ (green line)

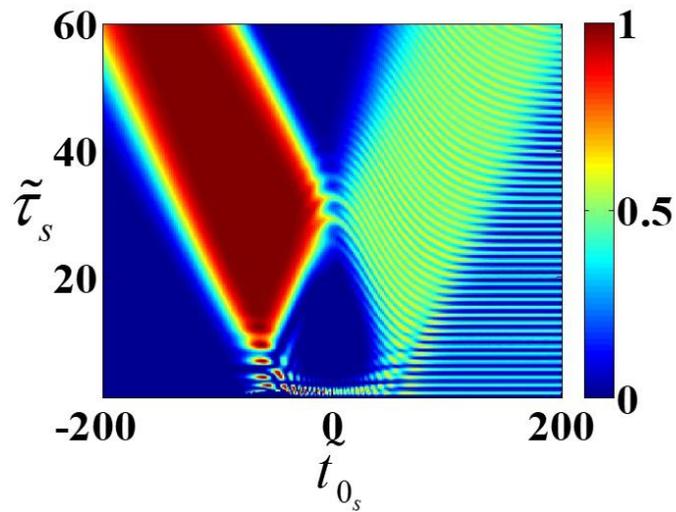

FIG. 4. FPE versus Stokes pulse width and pulse delay for $\tilde{\Delta}_{P,S} = 0, \tilde{\tau}_P = 30$ and $\tilde{\omega}_{\max_{P,S}} = 1$

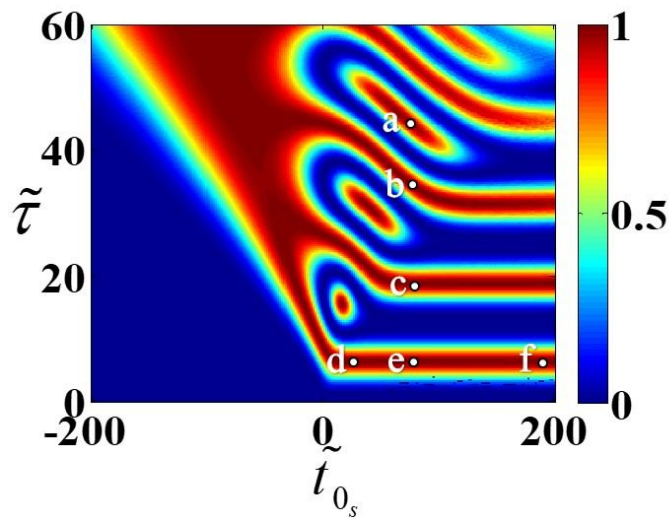

FIG. 5. FPE versus pulse width and pulse delay for $\tilde{\Delta}_{P,S} = 0$ and $\tilde{\omega}_{\max_{P,S}} = 0.1$

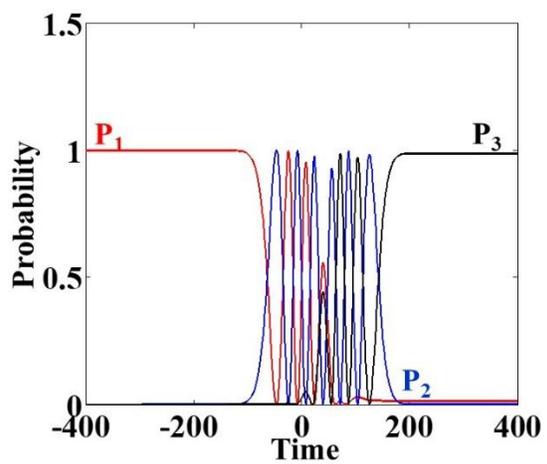

(a)

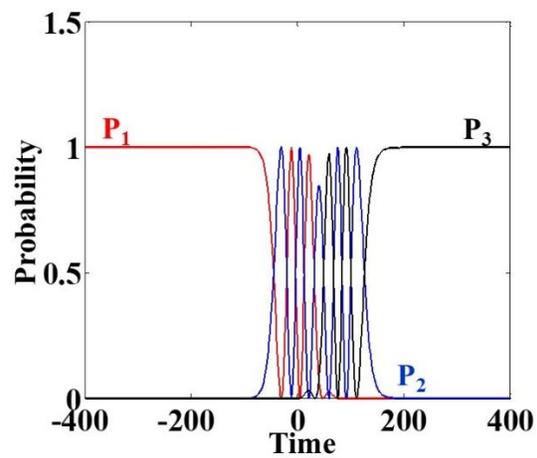

(b)

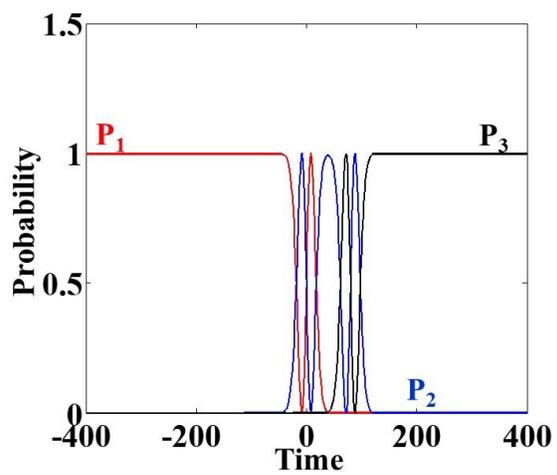

(c)

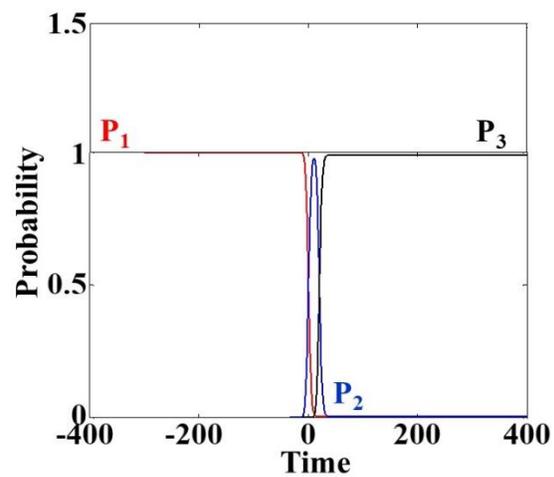

(d)

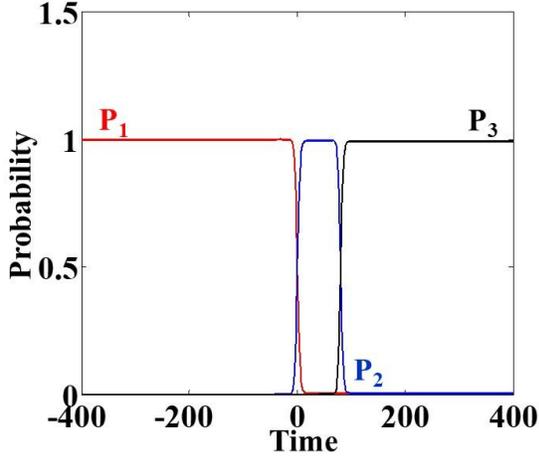 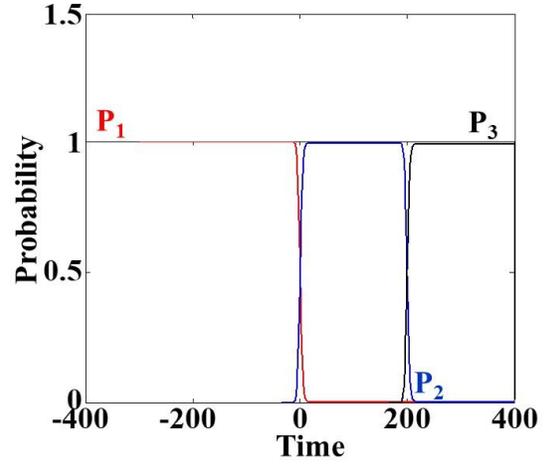

(e)                                  (f)

FIG. 6. Transition probabilities versus time for three states and $\tilde{\Delta}_{P,S} = 0$, $\tilde{\omega}_{\max_{P,S}} = 1$ (a) $\tilde{t}_{0_S} = 80$,

$\tilde{\tau} = 43.5$  (b) $\tilde{t}_{0_S} = 80$, $\tilde{\tau} = 34$  (c) $\tilde{t}_{0_S} = 80$, $\tilde{\tau} = 19$  (d) $\tilde{t}_{0_S} = 20$, $\tilde{\tau} = 6$  (e) $\tilde{t}_{0_S} = 80$, $\tilde{\tau} = 6$  (f)

$\tilde{t}_{0_S} = 200$, $\tilde{\tau} = 6$

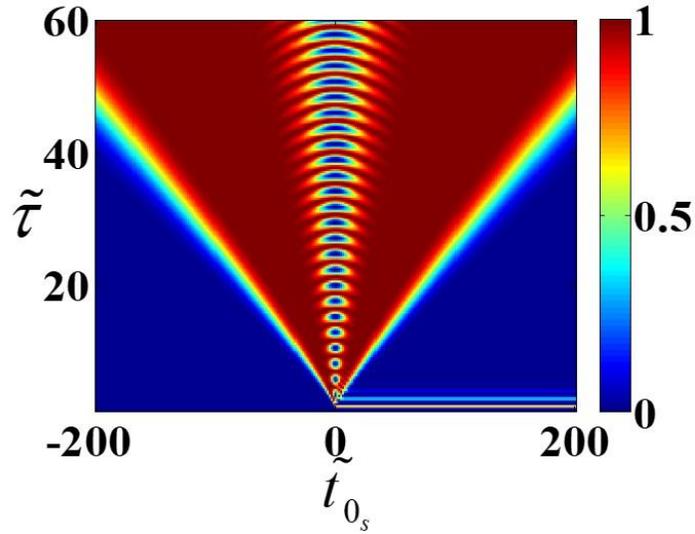

FIG. 7. FPE versus pulse width and pulse delay for $\tilde{\Delta}_{P,S} = 0.5$ and $\tilde{\omega}_{\max_{P,S}} = 1$

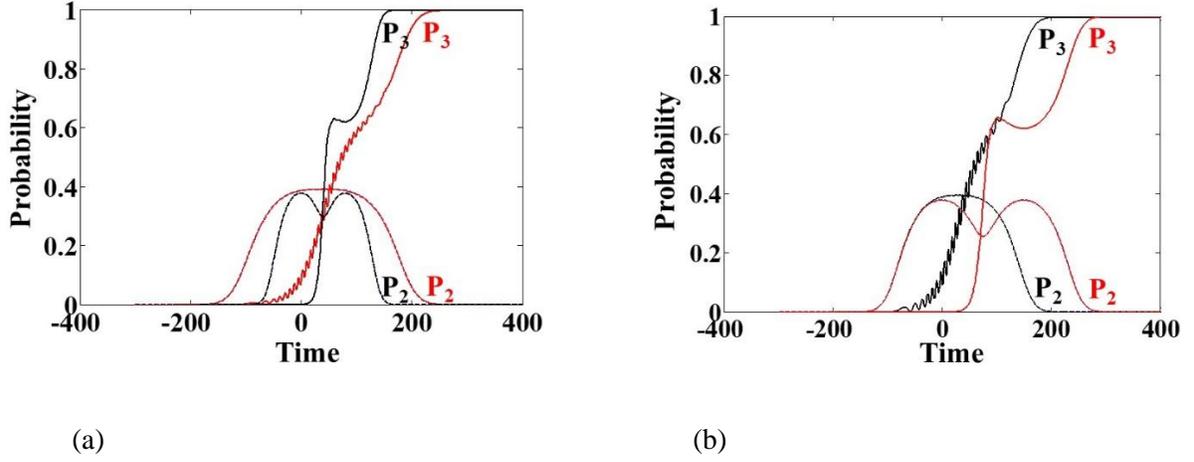

(a)                                        (b)

FIG. 8. Transition probabilities versus time for |2⟩ and |3⟩ states and $\tilde{\Delta}_{P,S} = 0.5, \tilde{\omega}_{\max_{P,S}} = 1$ (a)

$\tilde{t}_{0_S} = 80$, $\tilde{\tau} = 30$ (black line) and $\tilde{\tau} = 60$ (red line) (b) $\tilde{\tau} = 50$, $\tilde{t}_{0_S} = 60$ (black line) and

$\tilde{t}_{0_S} = 140$ (red line)

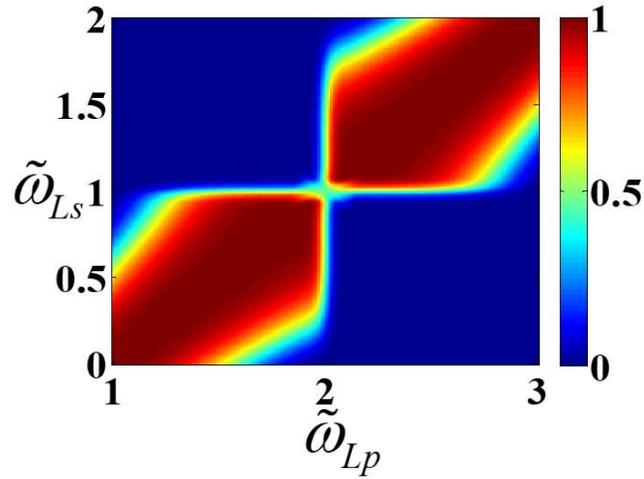

FIG. 9. FPE versus Stokes and pump laser frequencies for $\tilde{t}_{0_S} = 80, \tilde{\tau} = 50, \tilde{\omega}_{\max_{P,S}} = 1$

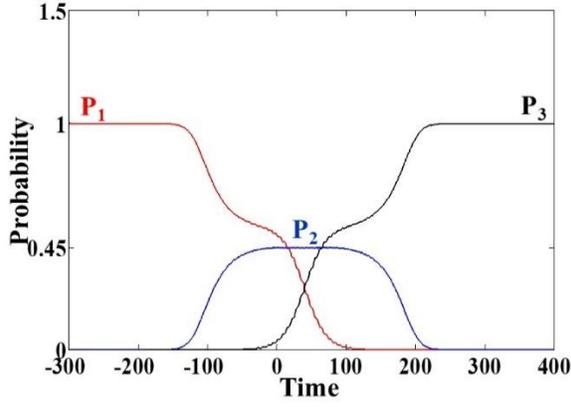

(a)

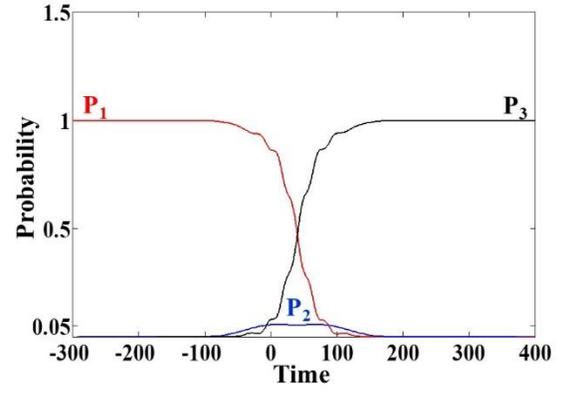

(b)

FIG. 10. Transition probabilities versus time for three states and $\tilde{t}_{0_S} = 80, \tilde{\tau} = 50, \tilde{\omega}_{\max_{P,S}} = 1$

(a) $\tilde{\Delta}_{P,S} = 0.2, \tilde{\omega}_{L_P} = 2.2, \tilde{\omega}_{L_S} = 1.2$ (b) $\tilde{\Delta}_{P,S} = 4, \tilde{\omega}_{L_P} = 6, \tilde{\omega}_{L_S} = 5$